\documentclass[useAMS,usenatbib]{mn2e}
\usepackage{graphicx}
\usepackage{amssymb}
\usepackage{amsmath}
\usepackage{color}

\title[Massive star formation in {galaxies with excess UV emission}]
  {Massive star formation in galaxies with excess UV emission}
\author[S. Erroz-Ferrer et al.]
{Santiago Erroz-Ferrer,$^{1,2}$\thanks{Email: serroz@iac.es} Johan H. Knapen,$^{1,2}$ Elena A. N. Mohd Noh Velast\'in,$^{3}$
\newauthor Jenna E. Ryon$^{4,7}$ and Lea M. Z. Hagen$^{5,6,7}$ \\
$^{1}${Instituto de Astrof\'isica de Canarias, V\'ia L\'actea s/n 38205 La Laguna, Spain}\\
$^{2}${Departamento de Astrof\'isica, Universidad de La Laguna, 38206 La Laguna, Spain}\\
$^{3}${School of Physics and Astronomy, University of St Andrews, North Haugh, Fife KY16 9SS St Andrews, UK}\\
$^{4}${Department of Astronomy, University of Wisconsin - Madison, 475 N. Charter St., Madison, WI, 53706, USA}\\
$^{5}${Department of Astronomy \& Astrophysics, The Pennsylvania State University, 525 Davey Laboratory, University Park, PA 16802, USA}\\
$^{6}${Institute for Gravitation and the Cosmos, The Pennsylvania State University, University Park, PA 16802, USA}\\
$^{7}${The Observatories of the Carnegie Institution for Science, 813 Santa Barbara Street, Pasadena, CA 91101, USA}}
\date{Accepted 2013 September 20.  Received 2013 August 28; in original form 2013 February 28}

\pagerange{\pageref{firstpage}--\pageref{lastpage}} \pubyear{2013}

\begin{document}

\label{firstpage}

\maketitle

\begin{abstract}
From an analysis of almost 2000 GALEX images of  galaxies with morphological types ranging from E to Sab, we have found a significant subset (28\%) that show UV emission outside $R_{25}$. We have obtained H$\alpha$ imaging of ten such galaxies, and find that their star formation rates are similar in the UV and in H$\alpha$, with values ranging from a few tenths to a few $M_{\odot} $ yr$ ^{-1} $. Probably because our sample selection is biased towards star-forming galaxies, these rates are comparable to those found in disk galaxies, although the star formation rates of the elliptical galaxies in our sample are well below $1\,M_{\odot} $ yr$ ^{-1}$. We confirm that the extended UV emission in our sample is caused by massive star formation in outer spiral arms and/or outer (pseudo) rings, rather than by alternative mechanisms such as the UV upturn.
\end{abstract}


\section{Introduction}
 \label{section1}

Galaxies with morphological types between E and Sab are generally characterised by low star formation rates (SFRs) (e.g., \citealt{Kennicutt1983}; \citealt{Trinchieri1991}; \citealt{Caldwell1991}; \citealt{Caldwell1994}; \citealt{Pogge1999}; \citealt{James2004}; \citealt{Hameed2005}) and low gas content (e.g., \citealt{Roberts1969} or \citealt{Roberts1994}). They  are common in the local Universe, and are thought to be evolved galaxies, formed either through a series of mergers of more or less gas-rich galaxies happening on cosmological timescales (primarily the ellipticals; see e.g., \citealt{Toomre1972}; \citealt{Schweizer1992}; \citealt{Kormendy2009}), or through a process of continued secular evolution which has slowly transformed later-type into earlier-type galaxies (primarily S0-type galaxies, see, e.g., \citealt{Laurikainen2010}). 

Such galaxies span a very wide range of SFRs: from near-zero to several $M _{\odot} $ yr$ ^{-1} $. Their SFRs are generally low ($<0.6 M _{\odot} $ yr$ ^{-1} $ in galaxies with T-types between 0 and 2), and their H$\alpha$ equivalent widths (EWs) decrease systematically from late-type galaxies to early-type spirals (\citealt{James2004}). This indicates that the latter have lower specific SFRs (or SFRs per unit stellar mass, as indicated through the EWs), and generally less massive star formation.

Several pieces of observational evidence have recently led to the realisation that ellipticals, S0s and early-type spirals (ETSs) may be much less relaxed and inactive than previously considered. \citet{Young1996} and \citet{Usui1998} identified numerous ETSs with SFRs comparable to the most prolifically star-forming late-type spirals. The SAURON survey has shown that the majority of E and S0 galaxies exhibit line emission from their central regions \citep{Sarzi2006}, although in at most half of these cases the emission is a direct result of young massive stars \citep{Sarzi2010}. 

Most of the SFR studies quoted so far were based on H$\alpha$ emission, but one can also consider another tracer, namely ultraviolet (UV) continuum emission. Young, massive stars emit most of their energy in this part of the spectrum, and the flux emitted in UV in spiral galaxies is an excellent measure of their current SFR (e.g. \citealt{Donas1987}; \citealt{Kennicutt1998}). Thanks to the \textit{Galaxy Evolution Explorer} (\textit{GALEX}; \citealt{Martin2005}), deep UV observations with moderately good spatial resolution for large numbers are now available for many nearby galaxies. 

The starting point of this project was the analysis of \textit{GALEX} UV images of ellipticals, S0s, and ETSs. In a significant fraction of these galaxies (see Sect. \ref{section2}), we found UV emission in the outer parts of the discs, in the form of complete or incomplete rings, spiral arms or arm fragments, or in more chaotic structures. Here, we investigate the origin of this emission by obtaining H$\alpha$ imaging of a few selected galaxies. These images will allow us to answer the important question of whether the excess UV emission is caused by massive star formation with enough OB stars to produce H$\alpha$ emission, or whether the initial mass function is not populated enough to produce significant amounts of the most massive stars that produce H$\alpha$ emission (see \citet{Lee2011}), a scenario which would lead to star formation causing UV, but no H$\alpha$ emission.

An alternative scenario that can, at least in principle, play a role here is the UV upturn phenomenon (\citealt{Dorman1995}), where UV radiation is emitted by old stellar populations, with as most likely source extreme horizontal branch stars. This phenomenon occurs in elliptical galaxies and bulge regions in spirals (\citealt{OConnell1999}), and it is unlikely to be an explanation for the UV emission we are concerned with here, originating outside $R_{25}$. Detection of H$\alpha$ emission in our images would all but exclude this possibility.  \citet{Salim2012} studied a sample of ellipticals, S0s, and ETSs and found galaxy-scale star formation in the form of UV rings of various sizes and morphologies, consistent with low-level SFRs of $\sim0.5 M _{\odot} $ yr$ ^{-1} $, but which they considered inconsistent with the UV upturn.

This paper aims to investigate which of these scenarios is causing the UV emission observed by {\it GALEX} in the sample galaxies. It is organized as follows: Section \ref{section2} gives a description of the sample, and Section \ref{section3} describes the observations and data reduction. The results are presented in Section \ref{section4},  and discussed in Section \ref{section5}. A summary of our conclusions is given in Section \ref{section6}.


\section{Target selection}
 \label{section2}

We have analysed 4822 images of large ($ > $ 0.8 arcmin) ellipticals, S0s, and ETSs from the \textit{GALEX} Large Galaxy Atlas (GLGA; \citealt{Seibert2007}). In that sample, 80 galaxies had no appreciable \textit{GALEX} detections, leaving 4742 classified galaxies. These galaxies cover the Hubble type range of E through Sab, and T-types from -5.0 to 2. Due to the fact that not all the galaxies were imaged at the same depth, we chose only galaxies that were probed to 27 mag arcsec$ ^{-2} $ at a signal-to-noise ratio of 5, leading to a sample of 1899 galaxies.

To perform the classification for extra UV emission in the galaxy sample, we compared optical data from the literature and \textit{GALEX} images for each of the 1899 galaxies. We visually inspected the galaxy regions outside $R_{25}$ in all images, looking for any clumpy morphological component, such as rings, spiral arms, or fragments of spiral arms, and for evidence of streams of material. We found that 18\% of E galaxies are in these categories, increasing to 28\% for S0 and 45\% for S0/a-ab. We also classified extended UV emission located beyond the optical $R_{25}$ radius that is not in the form of any specific feature, and found this in fractions ranging from 1\% in E to 3\% in S0 and 7\% in S0/a-Sab. Taking everything into account, we find UV emission on the outskirts of 531 of the 1899 galaxies in the sample (corresponding to 28\%).

From the sample of 531 galaxies with extra UV emission, we have selected ten galaxies to be observed in H$\alpha$, randomly selected at the time of the narrow-band observations, according to the visibility in the sky, but not selecting on properties like morphological type\footnote{We did ensure the presence of two elliptical galaxies in our observed sample.} or the presence of rings or spirals: NGC~160, NGC~262, NGC~4698, NGC~5173, NGC~5389, NGC~5982, NGC~6962, NGC~7371, NGC~7787 and PGC~065981. The general properties of the ten galaxies in the sample, ranging in morphological type from E to Sab, are presented in Table \ref{sample}.

\begin{table*}
\centering
\caption{General properties of the galaxies in the sample. Notes. (1) Morphological type and Hubble T-types from The Third Reference Catalogue of Bright Galaxies (RC3; \citealt{RC3}). (2)  Updated morphological classifications from \citealt{Buta2010} and R.J. Buta et al. 2013 (in prep), where "double stage" galaxies are allowed (i.e. large-scale S0 or S0/a galaxies with smaller-scale inner spirals). (3) Adopted values of the distances, calculated after applying the Virgo, GA and Shapley corrections, with \textit{H}$_0$= 73 $\pm$ 5 km s$^{-1}$ Mpc$^{-1}$, from the NASA/IPAC Extragalactic Database (NED). As the distances have been measured using the Tully-Fisher relationship, the uncertainties in the distances are estimated to be 20\% of the given value. (4) Apparent major isophotal diameter, \textit{D}$ _{25} $, measured at or reduced to the surface brightness level B = 25.0 mag arcsec$ ^{-2} $, as explained in Section 3.4.a, page 21, of Volume I of the printed RC3.}
 \label{sample}
\begin{tabular}{|c|c|c|c|c|c|c|}
\hline
\hline
Galaxy Name & Type (RC3) & Updated Type (Buta) & \textit{D} (Mpc) & \textit{D}$ _{25}$ (arcmin) \\
 & (1) &  (2) & (3) & (4) \\
 \hline
NGC~160    &   (R)SA0$^{\wedge}$+pec & & 70.5 $\pm$ 14.1 & 2.95 \\
NGC~262    &   SA0/a?(s)   &   & 60.7 $\pm$ 12.1 & 1.07  \\
NGC~4698   &   SA(s)ab    &(R)SA(rr)0/a/E2    & 13.7 $\pm$ 2.7 & 3.98  \\
NGC~5173   &   E0:     & E$ \wedge $+1    &  40.6 $\pm$ 8.1 & 1.78 \\
NGC~5389   &   SAB0/a?(r)   &              &  32.2 $\pm$ 6.4 & 3.47 \\
NGC~5982   &   E3   & E2     &  47.9 $\pm$ 9.6 & 2.57 \\
NGC~6962   &   SAB(r)ab              &               & 61.1 $\pm$ 12.2 & 2.88 \\
NGC~7371   &   (R)SA0/a?(r)          & S\underline{A}B(s)ab         & 37.9 $\pm$ 7.6 & 2.04  \\
NGC~7787   &   (R')SB0/a?(rs)        &          & 89.8 $\pm$ 18.0 & 1.78 \\
PGC~065981 &   SAB(s)ab?             &               & 122.5 $\pm$ 14.5 & 1.55 \\
\hline
\end{tabular}
\end{table*}


\section{Observations and data reduction}
 \label{section3}

\subsection{ALFOSC \textit{R}-band and H$\alpha$}
 \label{section31}

We have obtained \textit{R}-band and H$\alpha$ narrow band images with the Andalucia Faint Object Spectrograph and Camera (ALFOSC) at the 2.5 m Nordic Optical Telescope (NOT), in La Palma, during the nights of July 24th and 25th 2011, and June 2nd 2013. The ALFOSC field of view (FOV) is 6.4 arcmin$ ^{2} $, at a pixel scale of 0.19 arcsec. The \textit{R}-band images were taken using a Bessel filter with central wavelength 6500\AA{} and full width half maximum (FWHM) of 1300\AA{}. We used five different H$\alpha$ filters (central wavelength/FHWM, in \AA{}) depending on the galaxy's redshift (6655/55: NGC~160, NGC~262 and NGC~6962; 6615/50: NGC~5173, NGC~5389, NGC~5982 and NGC~7371; 6584/36: NGC~4698; 6704/45: NGC~7787; 6745/50: PGC~065981). The exposure times per galaxy were 5 $ \times $ 120 seconds for the \textit{R}-band images and 5 $ \times $ 900 seconds for the H$\alpha$ images. The typical seeing was around 1 arcsec, ranging from 0\farcs6 to 1\farcs6.

The \textit{R}-band and H$\alpha$ images have been reduced using {\sc iraf}. Firstly, bias and flat corrections were made. The sky was subtracted before the combination of the individual images. The \textit{R}-band images were photometrically calibrated using Landolt photometric standards. Foreground stars were removed manually from the images. Then, we performed surface photometry using the {\sc ellipse} task in {\sc iraf}. We checked visually the resulting ellipses in order to ensure that no localized star-forming regions were excluded from the integration. The fluxes were corrected for Galactic absorption, \textit{A(R)}, taken from NED, from the \citet{Schlafly2011} recalibration of the \citet{Schlegel1998} infrared-based dust map. We derived the instrumental magnitude of each galaxy, and computed the $ m_{R} $ magnitudes and the corresponding absolute $ R $ magnitudes ($M_{R}$) using the distances in Table \ref{sample}. The results for the whole galaxy measurements are presented in Table \ref{photometry}. The errors in the absolute $ R $ magnitudes mainly come from the uncertainty in the distance measurement (20\%).

The continuum was subtracted from the H$\alpha$ images following the procedures outlined in \citet{Knapen2004}, \citet{Bradley2006} and \citet{Sanchez-Gallego2012}. The images were photometrically calibrated using spectrophotometric standard stars. We derived the observed fluxes and computed the corresponding luminosities for the whole galaxies, calculated as
 
 \begin{equation}
 L({\rm H\alpha}) [{\rm erg/s}]=4\pi D^{2} (3.086 \times 10^{24})^{2}F_{{\rm H\alpha}}^{*},
\end{equation}  

with \textit{D} the distance to the galaxy in Mpc (Table \ref{sample}) and $F_{{\rm H\alpha}}^{*}$ the flux corrected for Galactic absorption.  As we are interested in the outer features (i.e., external rings or spiral arms), we isolated them and measured the fluxes and luminosities of these regions as well. These observed values were also corrected for Galactic absorption (see Section \ref{section42}).

\subsection{\textit{GALEX} UV}
 \label{section32}

We have obtained \textit{GALEX} images through The Barbara A. Mikulski Archive for Space Telescopes (MAST\footnote{http://archive.stsci.edu/}), using data from the All-sky Imaging Survey (AIS), Medium Imaging Survey (MIS), Guest Investigator Program (GIP) and Nearby Galaxy Survey (NGS). The scientific objectives and characteristics of \textit{GALEX}, as well as the surveys are described in \citet{Martin2005} and \citet{Morrissey2005}. The \textit{GALEX} FOV is 1.2 degrees and is circular. The effective wavelengths of the two channels (FUV and NUV) are 1516 and 2267 \AA{}, and the image resolution (FWHM) is 4\farcs3 and 5\farcs3, respectively, for the FUV and NUV channels.

We used the same regions defined in the H$\alpha$ images to perform the photometry in the FUV and NUV images, firstly for the whole galaxy and secondly for the specific features. We used the \textit{GALEX} zero point magnitudes presented in \citet{Morrissey2007} to convert intensities into magnitudes. The sky background was subtracted before performing the photometry. The FUV and NUV flux densities were corrected for Galactic absorption using the \citet{Schlafly2011} recalibration of the \citet{Schlegel1998} infrared-based dust map and the Galactic Extinction curve derived by \citet{Cardelli1989} for a total-to-selective extinction of $ R_{V} $=3.1. Specifically, $ A({\rm FUV})_{\rm MW} $ = 7.9 $ E(B-V) $ and $ A({\rm NUV})_{\rm MW}  $=~8.0~$ E(B-V) $. The foreground-extinction-corrected UV magnitudes are presented in Table \ref{photometry}, and the measurements analysed in Section \ref{section42}.

\begin{table*}
\caption{Results from the surface photometry. Column I) Galaxy name. Columns II, III and IV) Galactic absorption $ A(R) $, $ A(B) $ and $ A(V)$ from the \citet{Schlafly2011} recalibration of the \citet{Schlegel1998} dust map. Column V): Colour excess $ E(B-V) $. Column VI) Absolute \textit{R} magnitude. Columns VII and VIII) FUV and NUV asymptotic magnitudes. \textit{R} and UV magnitudes have been corrected for Galactic absorption. Column IX) (FUV - NUV) colour.}
 \label{photometry}
\centering
\begin{tabular}{|c|c|c|c|c|c|c|c|c|}
\hline
Galaxy name & $ A(R) $ & $ A(B) $ & $ A(V) $ & $ E(B-V) $ & $ M_{R} $ & FUV & NUV & FUV - NUV\\
 & (mag) & (mag) & (mag) & (mag) & (mag)& (mag) & (mag) & (mag)\\
\hline 
NGC~160 & 0.071 & 0.119&0.090&0.029 & $-$21.77 $\pm$ 1.00 & 17.31 $\pm$ 0.03 & 16.81 $\pm$ 0.03 &0.50 $\pm$ 0.05\\
NGC~262 & 0.145 & 0.242&0.183&0.059 & $-$20.27 $\pm$ 1.00 & 16.19 $\pm$ 0.04 & 15.90 $\pm$ 0.04 &0.29 $\pm$ 0.06\\
NGC~4698 & 0.056 &0.094&0.071& 0.023 & $-$20.28 $\pm$ 1.00 & 16.38 $\pm$ 0.03 & 15.36 $\pm$ 0.03 &1.01 $\pm$ 0.05\\
NGC~5173 & 0.059 &0.063&0.048& 0.015 & $-$20.61 $\pm$ 1.00 & 17.10 $\pm$ 0.03 & 16.59 $\pm$ 0.03 &0.52 $\pm$ 0.04\\
NGC~5389 & 0.043 &0.073&0.055& 0.018 & $-$20.93 $\pm$ 1.00 & 17.60 $\pm$ 0.03 & 16.96 $\pm$ 0.03 &0.64 $\pm$ 0.04\\
NGC~5982 & 0.038 &0.171&0.129& 0.042 & $-$22.65 $\pm$ 1.00 & 17.92 $\pm$ 0.03 & 16.46 $\pm$ 0.04& 1.47 $\pm$ 0.05\\
NGC~6962 & 0.212 &0.355&0.269& 0.086 & $-$22.48 $\pm$ 1.00 & 15.74 $\pm$ 0.05 & 15.30 $\pm$ 0.05 & 0.44 $\pm$ 0.07\\
NGC~7371 & 0.128 &0.213&0.161& 0.052 & $-$21.00 $\pm$ 1.00 & 15.42 $\pm$ 0.04 & 14.92 $\pm$ 0.04 & 0.50 $\pm$ 0.06\\
NGC~7787 & 0.082 &0.137&0.104& 0.033 & $-$20.90 $\pm$ 1.00 & 18.28 $\pm$ 0.03 & 17.62 $\pm$ 0.04 & 0.66 $\pm$ 0.05\\
PGC~065981 & 0.102 &0.171&0.129& 0.042 & $-$22.27 $\pm$ 1.00 & 16.70 $\pm$ 0.04 & 16.33 $\pm$ 0.04 &  0.37 $\pm$ 0.05\\
  
\hline  
\end{tabular}
\end{table*}


\section{Results}
 \label{section4}

\subsection{Morphology: General}
\label{generalmorph}

We have found several common characteristics when comparing the morphology of the galaxies in the \textit{GALEX} UV and H$\alpha$ narrow band images, as presented in Figs. \ref{fig1}-\ref{fig3}. From the 531 galaxies in our sample  of galaxies with extra UV emission, we selected ten, and most of these have been classified as having rings or spirals. This can be confirmed in our UV and H$\alpha$ images. In all of these cases, the H$\alpha$ emission outlines the rings or the spiral arms, whereas the rings in the UV images can be easily recognised. In the \textit{R}-band images, however, the structures are not traced that well. There are three galaxies where the morphological classifications given by the RC3 are uncertain and which have not been studied by \citet{Buta2010}. In those cases, we have analysed our images and we can remove the uncertainties by confirming the RC3 classification (i.e., NGC~262:~SA(s)0/a, NGC~5389:~SAB(r)0/a and PGC~065981:~SAB(s)ab).

It is important to note the different angular resolution of the H$\alpha$ (seeing limited, with values from 0\farcs7 to 1\farcs6), and the \textit{GALEX} images (4\farcs3 and 5\farcs3 for the NUV and FUV images, respectively). As a consequence, the structures of the galaxies can look extended in the UV images as the emission is smeared out. Furthermore, as the UV traces older stars than H$\alpha$, the UV emission is expected to be more extended than the H$\alpha$. 

\subsection{Morphology: Individual galaxies}
 \label{section42}

\subsubsection{NGC~160}

NGC~160 has been classified as (R)SA0$^{\wedge}$+pec in the RC3. In the H$\alpha$ and UV images (panels a and c of Fig. \ref{fig1}), the outer ring is clearly identified. This is one of the best cases of correlation between H$\alpha$ and UV. The H$\alpha$ emission is more patchy than the UV in this ring. UV emission is detected in the centre, but H$\alpha$ emission from there is completely absent. The continuum-subtraction residuals in the nuclear region result from the large \textit{R}-band emission there, typical of a lenticular galaxy.

\subsubsection{NGC~262}
Morphologically, NGC~262 is classified as SA0/a?(s) in the RC3. In terms of nuclear activity, this galaxy has been classified as a Seyfert 2 galaxy, for the first time in the Second Reference Catalogue of Bright Galaxies (RC2; \citealt{RC2}). The nuclear emission is indeed strong, and dominates the flux of the galaxy. However, the spiral arms appear perfectly defined in both the H$\alpha$ (panel d in Fig. \ref{fig1}) and UV (panel f in Fig. \ref{fig1}) images.

\subsubsection{NGC~4698}
\begin{figure*}
\begin{center}
\includegraphics[scale=0.93]{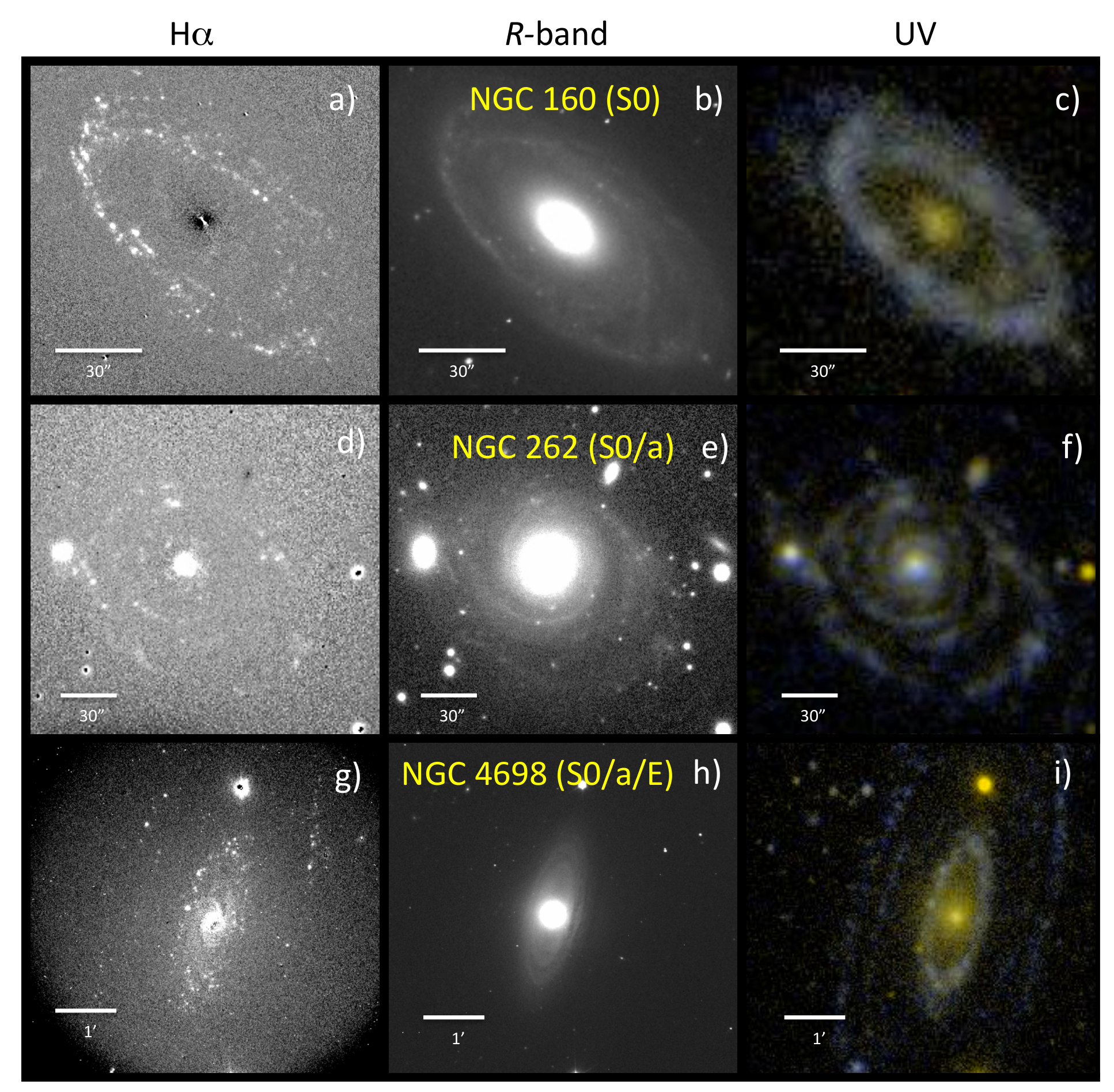}
\caption{Images on the left are H$\alpha$ continuum-subtracted images of the galaxies, those in the middle correspond to the \textit{R}-band images, and those on the right are false-color \textit{GALEX} maps of the galaxies. The RGB images used are `asinh' scaling versions \citep{Lupton2004} of the 2-pixel-smoothed FUV image (blue), the original NUV image (red), and a linear combination of the two (green). a), b) and c) NGC~160. d), e) and f) NGC~262. g), h) and i) NGC~4698. In the images, North is up and East to the left.}
\label{fig1}
\end{center}
\end{figure*}

This galaxy was classified in the RC3 as SA(s)ab. R. J. Buta et al. 2013 (in prep) classify the galaxy as (R)SA(rr)0/a/E2, a double-stage spiral, described by \citep{Vorontsov-Velyaminov1987} as cases where an inner spiral pattern gives a different type than the outer spiral pattern. In this case, we find a E-like bulge where the spiral arms arise. In The Carnegie Atlas of Galaxies (\citealt{Sandage1994}), the galaxy was classified in the earliest one-third of the Sa group because of the smooth inner disk, the large bulge with no recent star formation, and the tightly wound spiral arms, defined primarily by the dust which forms fragmentary lanes. This galaxy presents two outer rings visible in both the UV and H$\alpha$. There is a third not-so-well-defined ring in between the other two, which can hardly be seen in H$\alpha$ (panel g in Fig. \ref{fig1}) but which can be recognised in the UV image (panel i in Fig. \ref{fig1}). Here, as in NGC~160, the H$\alpha$ emission is more patchy than the UV emission. NGC~4698 is a low-luminosity Seyfert 2 galaxy (\citealt{Ho1995}), and the nuclear emission is therefore strong.

\subsubsection{NGC~5173}

NGC~5173 has been classified as an elliptical galaxy, which is perfectly recognizable in the \textit{R}-band image. However, in both the H$\alpha$ and UV wavelengths, some non-spherical emission can easily be seen. There are clumps of material emitting in H$\alpha$ and UV, tracing young stars and, therefore, star formation. This clumpy emission seems to be part of incomplete spiral arms, better traced in the H$\alpha$ image (Fig. \ref{fig2}a). The emission from the centre may not be due to massive star formation and can be related to the UV-upturn phenomenon. As a consequence, the derived SFRs (Sect. \ref{sectionSFRs}) may be overestimated for the central part, and for the whole galaxy.

\subsubsection{NGC~5389}
\begin{figure*}
\begin{center}
\includegraphics[scale=0.93]{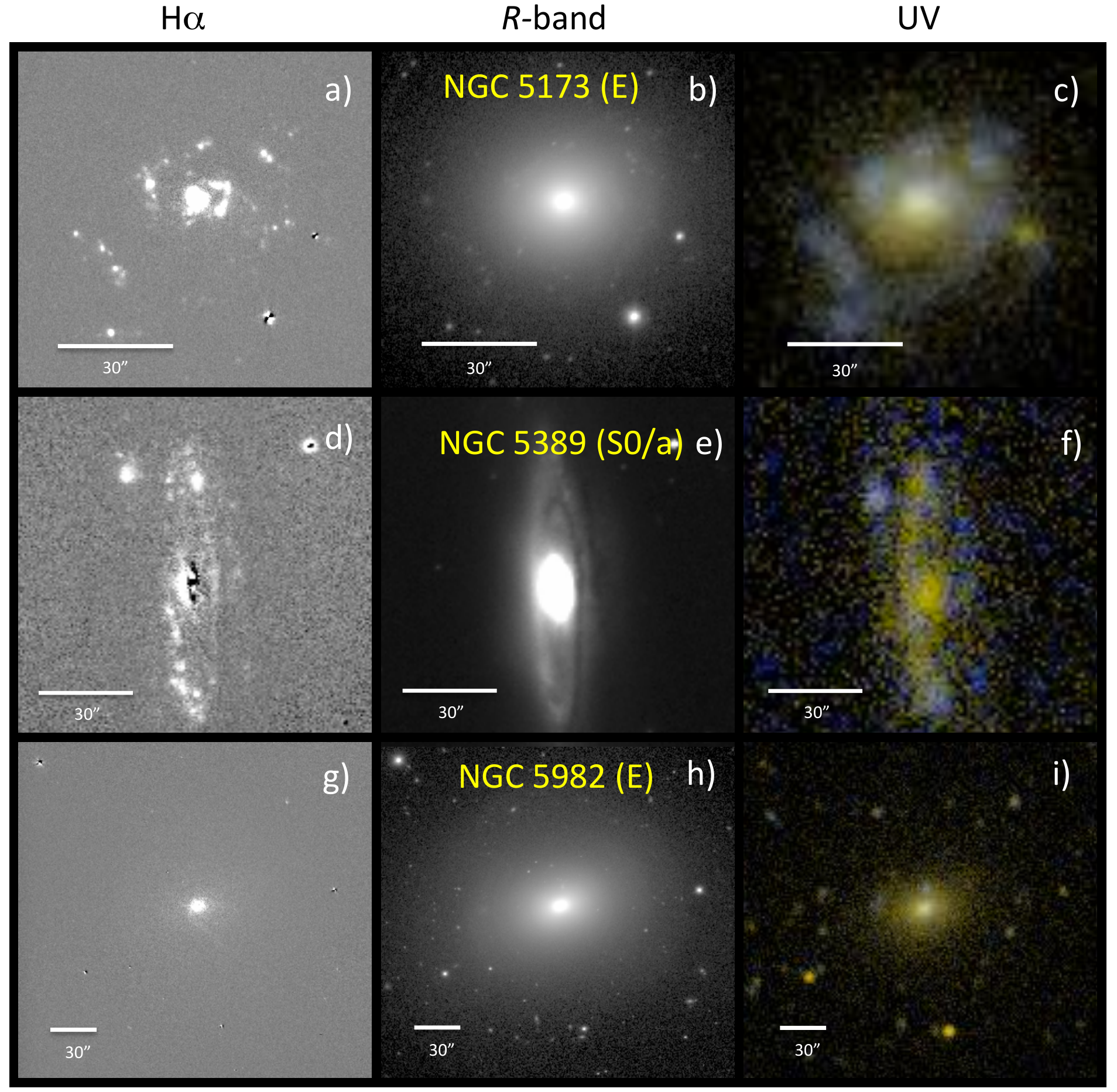}
 \caption{As Fig. \ref{fig1} but now: a), b) and c) NGC~5173. d), e) and f) NGC~5389. g), h) and i) NGC~5982. In the images, North is up and East to the left.} 
\label{fig2}
\end{center}
\end{figure*}

NGC~5389 is classified as SAB0/a?(r) in the RC3. The ring can be identified in both H$\alpha$ and UV images (panels d and e of Fig. \ref{fig2}). However, in the UV image, the right part (East) of the ring is not as bright as the left one. This effect, also seen in the H$\alpha$ image but not as clearly as in the UV, can be due to the inclination of the galaxy, which can cause one part of the galaxy to be affected more by dust than the other.

\subsubsection{NGC~5982}

This galaxy has been classified as elliptical. In the \textit{R}-band, the galaxy shows the typical spherical shape of an elliptical. However, in both the H$\alpha$ an UV images, the central part shows some emission. In contrast to the other pure elliptical, NGC~5173, NGC~5982 does not seem to have any non-nuclear H$\alpha$ emission. As in the case of NGC~5173, the emission in the central parts may not be due to star formation, especially since NGC~5982 has been classified as a LINER (\citealt{Ho1997}). There may also be a contribution from the UV-upturn phenomenon. In the UV image, there are some blue blobs tracing the formation of young stars. 

\subsubsection{NGC~6962}
NGC~6962 is a SAB(r)ab galaxy. This galaxy presents two main, faint outer arms, easily recognised in both H$\alpha$ and UV images  (panels a and c of Fig. \ref{fig3}). As in the other cases, the H$\alpha$ emission is patchy, whereas the UV emission is more regular.

\subsubsection{NGC~7371}
NGC~7371, whose classification is (R)SA0/a?(r) in the RC3, shows hardly any emission from the inner part of the disc, which is devoid of recent star formation. \citet{Buta2010} revised the RC3 classification, changing it to S\underline{A}B(s)ab. The high spatial resolution of the H$\alpha$  (panel d in Fig. \ref{fig3}) and the \textit{R}-band (panel e in Fig. \ref{fig3}) images allows us to confirm the presence of the spiral arms, whereas in the UV image  (panel f in Fig. \ref{fig3}) the poor \textit{GALEX} spatial resolution leads to the false perception that the galaxy has a ring instead of spiral arms.

\subsubsection{NGC~7787}
NGC~7787 has been clasified as (R')SB0/a?(rs). In the H$\alpha$ and UV images (panels g and i of Fig. \ref{fig3}), we can distinguish parts of an outer (pseudo-)ring.

\subsubsection{PGC~065981}
\begin{figure*}
\begin{center}
\includegraphics[scale=0.93]{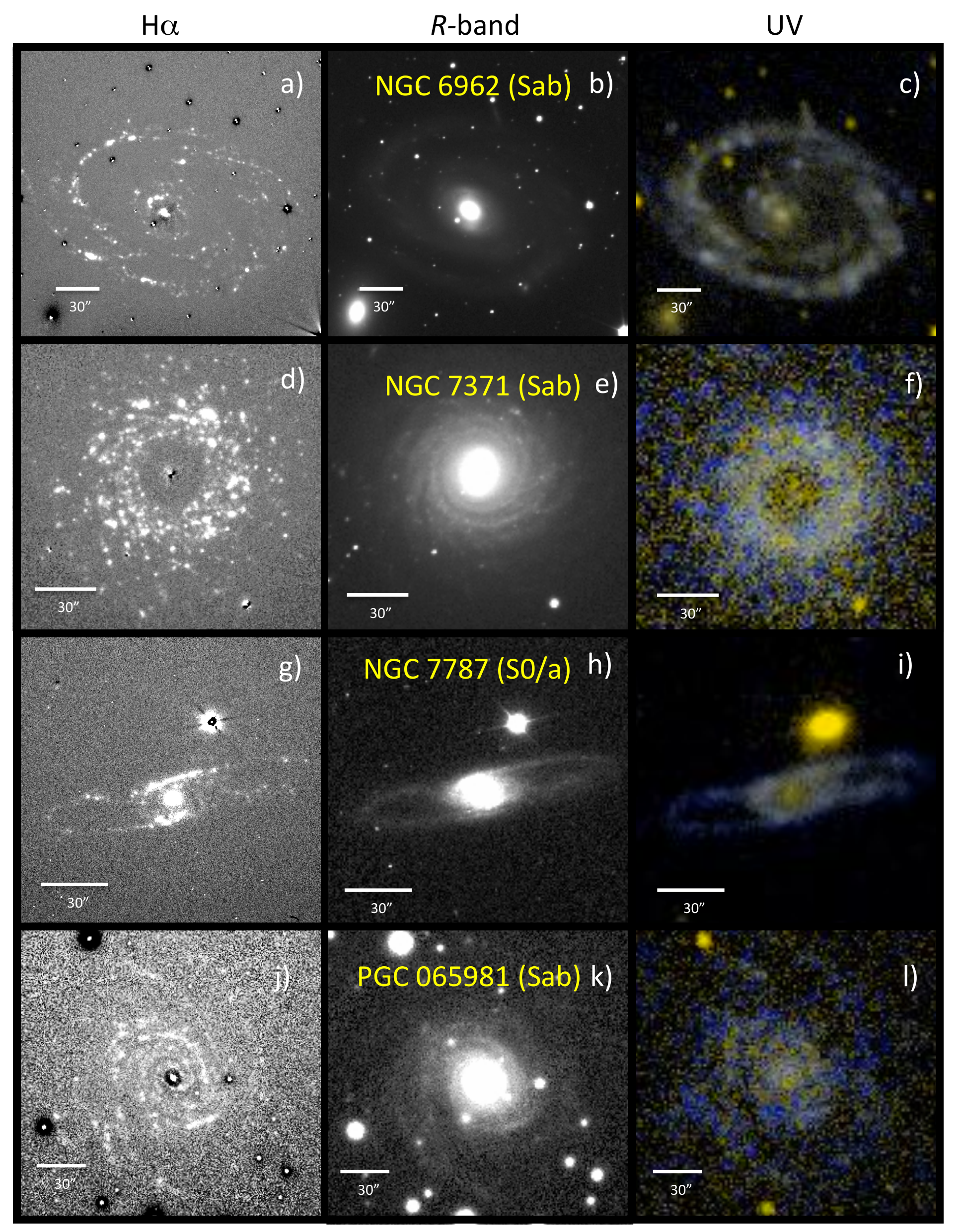}
 \caption{As Fig. \ref{fig1} but now: a), b) and c) NGC~6962. d), e) and f) NGC~7371. g), h) and i) NGC~7787. j), k) and l) PGC~065981. In the images, North is up and East to the left.}
\label{fig3}
\end{center}
\end{figure*}

 This galaxy, which has been classified as SAB(s)ab?, is a peculiar case. In both H$\alpha$ and UV images (panels j and l of Fig. \ref{fig3}), the spiral arms can be identified, more easily in the H$\alpha$ than in the UV image due to the different resolutions of the images or due to the different stellar populations (see Section \ref{generalmorph}). Besides, in the UV, the inter-arm zone cannot be easily defined.

\subsection{Star Formation Rates}
\label{sectionSFRs}

\subsubsection{Global measurements}

Several calibrations of SFR indicators have been presented in the literature in the past decades, using tracers across the electromagnetic spectrum: from the radio, infrared (IR), optical, UV or X-ray. Two of the most recent reviews on the subject are those by \citet{Kennicutt2012} and by \citet{Calzetti2012WS}. The availability of new observations in the past decade has led to a better determination of the uncertainties in the measurements of SFRs in galaxies, and the inclusion of new mixed indicators based on multi-wavelength data has reduced the uncertainties. The systematic errors have also been reduced with these techniques, particularly those related to dust attenuation, although uncertainties in the IMF remain as a limiting factor. We have to make assumptions about the underlying stellar IMF, expressed generally as a power-law relation (Salpeter 1995), double power law (Kroupa 2001) or log-normal distribution (Chabrier 2003). In this paper, we have adopted the IMF derived by Kroupa (2001), with a stellar mass range between 0.1-100 $ M_{\odot} $, and timescales $ t\geq 100 $ Myr for the UV and $ t\geq 6 $ Myr for H$\alpha$.

It is necessary to take into account both dust-obscured and dust-unobscured star formation, in other words, the SFR can be derived from direct stellar light and dust-processed light. There is a large list of star formation tracers: the direct stellar light emerging from galaxies can be quantified using stellar continuum UV and ionizing photon tracers (i.e., Hydrogen recombination lines such as H$\alpha$ or P$\alpha$ and forbidden lines like [O{\sc ii]$ _{3727} $}); whereas the dust-processed light can be traced using integrated IR [total-IR (TIR), far-IR (FIR)] or monochromatic IR bands (8, 24, 70 $ \mu$m and many more).

The launch of \textit{GALEX} has enabled the use of UV data, one of the most direct tracers of star formation, for large samples of galaxies. The primary drawback of the UV is the sensitivity to dust attenuation. The availability of IR maps and luminosities of nearby galaxies in the literature has allowed us to use TIR as a SFR calibrator to correct our UV measurements for this dust attenuation. Specifically, we have found IR data from the \textit{Infrared} \textit{Astronomical} \textit{Satellite} (\textit{IRAS}) for nine of the ten galaxies in our sample, and we have computed the TIR luminosity using data from the 25, 60 and 100 $ \mu$m \textit{IRAS} bands following the expression in  \citet{Dale2002}:

 \begin{align}
 L({\rm TIR}) = 2.403\nu L_{\nu}(25\mu \rm m) - 0.2454\nu L_{\nu}(60\mu \rm m) \nonumber \\
 + 1.6381\nu L_{\nu}(100\mu \rm m).
 \end{align}

These \textit{IRAS} values are collected in Table \ref{IRAS}. For PGC~065981, we have estimated $L$(TIR) from the FUV-NUV colour (\citealt{Cortese2006}):

\begin{equation}
 \label{beta}
\log\left[ \dfrac{{L(\rm TIR)}}{{L(\rm FUV)}}\right]=0.7 [2.201 (FUV - NUV) -1.804] + 1.3.
\end{equation}

\begin{table*}
\caption{IR photometry data.  Column I) Galaxy name.  Column II) 2MASS \textit{H}-band total magnitude. Columns III-VII) \textit{IRAS} flux densities for the 25, 60 and 100 $ \mu $m bands and the references, as well as the derived $ L ({\rm TIR}) $ using the expression given by \citet{Dale2002}. References: (1) NED (2) \textit{IRAS} Faint Source Catalogue, version 2.0 (\citealt{IRAS1990}). (3) \textit{IRAS} Faint Source Reject Catalog (\citealt{Moshir1992}). For PGC~065981, we have estimated $ L ({\rm TIR}) $ from the TIR/FUV ratio (\citealt{Cortese2006}).}
 \label{IRAS}
\begin{tabular}{|c|c|c|c|c|c|c|c|c|}
\hline
Galaxy name & \textit{H} & $ f_{\nu}(25) $ & $ f_{\nu}(60) $ & $ f_{\nu}(100) $ & Ref. & $ L({\rm TIR}) $ \\
& (mag) & (Jy) & (Jy) & (Jy) & & (erg s$ ^{-1} $) \\
\hline 
NGC~160 &9.738 & 0.034 $\pm$ 0.034 & 0.140 $\pm$ 0.048 & 0.650 $\pm$ 0.082 & (1) & (2.38 $\pm$ 1.15)$ \times $10$^{43}$\\
NGC~262 & 10.554&0.835 $\pm$ 0.025 & 1.290 $\pm$ 0.116 & 1.549 $\pm$ 0.201 & (2) & (1.33 $\pm$ 0.54)$ \times $10$^{44}$\\
NGC~4698 & 7.780&0.154 $\pm$ 0.154 & 0.258 $\pm$ 0.062 & 1.864 $\pm$ 0.168 & (2) & (2.99 $\pm$ 1.58)$ \times $10$^{42}$\\
NGC~5173 &10.258 & 0.031 $\pm$ 0.031& 0.350 $\pm$ 0.042 & 0.530 $\pm$ 0.159 & (1) & (6.05 $\pm$ 3.40)$ \times $10$^{42}$\\
NGC~5389 &8.883 & 0.014 $\pm$ 0.014 & 0.410 $\pm$ 0.018 & 1.960 $\pm$ 0.072 & (1) & (1.18 $\pm$ 0.49)$ \times $10$^{43}$\\
NGC~5982 &8.362 & 0.022 $\pm$ 0.022& 0.033 $\pm$ 0.033 & 0.370 $\pm$ 0.035 & (1) & (6.62 $\pm$ 3.24)$ \times $10$^{42}$\\
NGC~6962 &9.063 & 0.144 $\pm$ 0.144 & 0.338 $\pm$ 0.047 & 2.250 $\pm$ 0.383 & (2) & (6.61 $\pm$ 3.37)$ \times $10$^{43}$\\
NGC~7371 &9.508& 0.184 $\pm$ 0.184 & 1.104 $\pm$ 0.341 & 2.679 $\pm$ 0.311 & (3) & (2.94 $\pm$ 1.53)$ \times $10$^{43}$\\
NGC~7787 &11.327& 0.177 $\pm$ 0.177 & 0.641 $\pm$ 0.058 & 1.846 $\pm$ 0.185 & (2) & (1.29 $\pm$ 0.73)$ \times $10$^{44}$\\
 PGC~065981&11.226& - & - & - & - & (1.10 $\pm$  0.87)$ \times $10$^{44}$ \\ 
\hline  
\end{tabular}
\end{table*} 

We then used the empirical calibration factors from \citet{Kennicutt2009} and \citet{Hao2011} to compute the luminosities corrected for dust attenuation:

\begin{equation}
 \label{Lcorr1}
L({\rm FUV})_{\rm corr} = L({\rm FUV})_{\rm obs} + 0.46 L({\rm TIR}),
\end{equation}
\begin{equation}
 \label{Lcorr2}
L({\rm NUV})_{\rm corr} = L({\rm NUV})_{\rm obs} + 0.27 L({\rm TIR})\ {\rm and}
\end{equation}
\begin{equation}
 \label{Lcorr3}
L({\rm H\alpha})_{\rm corr} = L({\rm H\alpha})_{\rm obs} + 0.0024 L({\rm TIR}).
\end{equation}

There is one systematic error that can affect the use of these IR-based recipes in low-metallicity and other dust-free galaxies or in galaxies with low specific SFR, which might be our case in our sample galaxies: evolved stars can contribute significantly to the dust heating, causing the IR luminosity to be overestimated, as well as the corrected SFRs (see \citealt{Kennicutt2012} and references therein). As a matter of fact, the calculated SFRs derived from luminosities corrected using equations \ref{Lcorr1}, \ref{Lcorr2} and \ref{Lcorr3} should be understood as upper limits, so the SFRs will be between the dust-uncorrected values and the dust-corrected values using this IR-based recipe.

To avoid these possible systematic errors, \citet{Cortese2008} presented empirical relations from which the UV attenuation $A(\rm FUV)$ can be estimated taking into account the age-dependence of the TIR/FUV versus $A$(FUV) relation. They adopt the formalism used by \citet{Gavazzi2002}, where $ \tau_{\rm age} $ is the time at which the SFR reaches the highest value over the whole galaxy (so it should not be understood as the real age of the stellar populations). Thus, a short $ \tau_{\rm age} $ corresponds to galaxies dominated by old stellar populations while long $ \tau_{\rm age} $ corresponds to young galaxies (i.e., star-forming). Cortese et al. recommend finding first a good proxy for $ \tau_{\rm age} $ using the FUV - $ H $ colour:

 \begin{equation}
  \label{taus}
 \log(\tau_{\rm age})=-0.068({\rm FUV} - H) + 1.13
 \end{equation}

and afterwards calculating $A$(FUV) or $A$(NUV) from their polynomial fit.

On the other hand, there have been several proposals in the literature to correct H$\alpha$ luminosities for internal dust extinction. Commonly used is the constant value of $ A({\rm H}\alpha) = 1.1 $ mag (\citealt{Kennicutt1983}). However, this value can change depending on several factors as morphological type or inclination (e.g., \citealt{James2005}). Here, we are using the absolute $ R $ magnitudes of the galaxies ($ M_{R} $) from Table \ref{photometry} to calculate $A$(H$\alpha$) as proposed by \citet{Helmboldt2004}:

\begin{equation}
  \label{helmboldt}
 \log[A({\rm H}\alpha)]= (-0.12 \pm 0.048) M_{R} + (-2.5 \pm 0.96)
 \end{equation}

The resulting values for the dust extinction corrections using TIR, and the methods of \citet{Cortese2008}, and \citet{Helmboldt2004} have been collected in Table \ref{extinction}. In Section \ref{section5} we will discuss the advantages and drawbacks of using the different methods.

\begin{table*}
\caption{  Column I) Galaxy name. Column II) $ \tau_{age} $ derived from eq. \ref{taus}. Column III) Extinction coefficient for H$\alpha$ using L(TIR). Column IV) Extinction coefficient for H$\alpha$ using eq. \ref{helmboldt}. Columns V-VI) Extinction coefficients for FUV using L(TIR) and \citet{Cortese2008} to correct for internal dust extinction. Columns VII-VIII)The same as V-VI but for NUV.}
 \label{extinction}
\begin{tabular}{|c|c|c|c|c|c|c|c|c|c|}
\hline
   Galaxy name &  $ \tau_{age} $&  $A({\rm H}\alpha)_{\rm TIR} $ &  $A({\rm H}\alpha)_{\rm R} $ &  $A({\rm FUV})_{\rm TIR} $ &  $A({\rm FUV})_{\rm Cort} $ &  $A({\rm NUV})_{\rm TIR} $ &  $A({\rm NUV})_{\rm Cort} $\\
  &  (Gyr) &  (mag)  &  (mag)  &  (mag)  &  (mag)  &  (mag)  &  (mag)\\
\hline
 NGC 160  &  4.0 &  0.60 &  1.30 &  1.24 &  1.45 &  0.82 &  1.09\\ 
 NGC 262  &  5.4 &  1.09 &  0.86 &  2.07 &  1.93 &  1.68 &  1.63\\
 NGC 4698&  3.6 &  0.58 &  0.86 &  1.51 &  0.91 &  0.75 &  0.80\\
 NGC 5173&  4.6 &  0.31 &  0.94 &  0.93 &  1.70 &  0.58 &  1.46\\
 NGC 5389&  3.4&  0.74 &  1.03 &  2.22 &  0.69 &  1.53 &  0.51\\
 NGC 5982&  3.2&  0.15 &  1.65 &  1.29 &  0.36 &  0.44 &  0.52\\
 NGC 6962&  4.6&  0.70 &  1.58 &  1.14 &  1.73 &  0.78 &  1.40\\
 NGC 7371 &  5.4 & 0.51 &  1.05 &  1.04 &  1.95 &  0.67 &  1.66\\
 NGC 7787 &  4.6 & 1.12 &  1.02 &  3.17 &  1.56 &  2.39 &  1.33\\
 PGC 065981 &  5.8 &  1.50 &  1.49 &  1.15 &  1.15 &  0.81 &  0.87\\
\hline  
\end{tabular}
\end{table*}

The dust-corrected luminosities can be now used to derive dust-corrected SFR measurements for the whole galaxies, using the expressions in \citet{Kennicutt2009} and \citet{Hao2011}:


 \begin{equation}
  \label{SFR1}
{\rm SFR}{\rm (FUV)} (M_{\odot}  {\rm yr^{-1}}) = 4.6 \times 10^{-44} L(\rm FUV),
 \end{equation}
 \begin{equation}
  \label{SFR2}
{\rm SFR}{\rm (NUV)} (M_{\odot}  {\rm yr^{-1}}) = 6.8 \times 10^{-44} L(\rm NUV)\ {\rm and}
 \end{equation}
  \begin{equation}
 \label{SFR3}
{\rm SFR}({\rm H\alpha}) (M_{\odot}  {\rm yr^{-1}}) = 5.5 \times 10^{-42} L({\rm H\alpha})
 \end{equation} 
 
The resulting SFRs can be found in Table \ref{SFRs2}. Apart from the errors intrinsically linked to the different methods, there are several factors to take into account when estimating the uncertainties in the measurements of the SFR (such as uncertainties related to the quality of the image, to the image reduction processes, or to the distance; see \citealt{Erroz2012} for a detailed discussion).

\begin{table*}
\caption{Global SFRs of the galaxies derived from the observed luminosities (i.e. not corrected for internal dust attenuation).}

 \label{SFRs}
\begin{tabular}{|c|c|c|c|}

\hline
Galaxy name   & SFR(H$\alpha$)$ _{\rm obs} $  & SFR(FUV)$ _{\rm obs} $  & SFR(NUV)$ _{\rm obs} $  \\
 & ($M _{\odot} $ yr$ ^{-1} $)  & ($M _{\odot} $ yr$ ^{-1} $) & ($M _{\odot} $ yr$ ^{-1} $) \\
\hline 
NGC~160  & 0.43 $\pm$ 0.17 & 0.24 $\pm$ 0.09 & 0.39 $\pm$ 0.15 \\
NGC~262  & 1.01 $\pm$ 0.41 & 0.49 $\pm$ 0.19 & 0.66 $\pm$ 0.26 \\
NGC~4698 & 0.06 $\pm$ 0.02 & 0.02 $\pm$ 0.01 & 0.06 $\pm$ 0.02 \\
NGC~5173 & 0.24 $\pm$ 0.10 & 0.09 $\pm$ 0.01 & 0.16 $\pm$ 0.06 \\
NGC~5389 & 0.16 $\pm$ 0.06 & 0.04 $\pm$ 0.01 & 0.07 $\pm$ 0.03 \\
NGC~5982 & 0.59 $\pm$ 0.24 & 0.06 $\pm$ 0.02 & 0.25 $\pm$ 0.10 \\
NGC~6962 & 0.96 $\pm$ 0.38 & 0.75 $\pm$ 0.30 & 1.16 $\pm$ 0.47 \\
NGC~7371 & 0.64 $\pm$ 0.26 & 0.39 $\pm$ 0.16 & 0.63 $\pm$ 0.25 \\
NGC~7787 & 0.94 $\pm$ 0.38 & 0.16 $\pm$ 0.06 & 0.30 $\pm$ 0.12\\
PGC~065981&0.49 $\pm$ 0.19 & 1.24 $\pm$ 0.49 & 1.81 $\pm$ 0.72 \\
  
\hline 
\end{tabular}

\end{table*}

\begin{table*}
\caption{Global SFRs of the galaxies. We present the measurements derived from the luminosities corrected for dust extinction using the extinction coefficients presented in Table \ref{extinction}. On the left, we present the resulting SFRs corrected for internal absorption using our preferred correction methods:  for H$ \alpha $, the method using the \textit{R}-band absolute magnitude presented in \citet{Helmboldt2004}; for FUV and NUV, the method using the FUV/TIR relationship in \citet{Cortese2008}. On the right, the alternative correction methods: for H$ \alpha $ only, the constant value of $ A({\rm H}\alpha) = 1.1 $ mag (\citealt{Kennicutt1983}); and for H$ \alpha $, FUV and NUV, the TIR correction using the empirical calibration factors from \citet{Kennicutt2009} and \citet{Hao2011}.}

 \label{SFRs2}
\begin{tabular}{|c|c|c|c|c|c|c|c|c|} 
\hline
& \multicolumn{3}{|c|}{Preferred corrections}& & \multicolumn{4}{|c|}{Alternative corrections}   \\ \cline{2-4} \cline{6-9}  
Galaxy name &  SFR(H$\alpha)_{\rm R} $ &  SFR(FUV)$_{\rm Cort} $ &  SFR(NUV)$_{\rm Cort} $ & &  SFR(H$\alpha)_{1.1} $ &  SFR(H$\alpha)_{\rm TIR} $ &  SFR(FUV)$_{\rm TIR} $ &  SFR(NUV)$_{\rm TIR} $ \\
&  ($M _{\odot} $ yr$ ^{-1} $)  & ($M _{\odot} $ yr$ ^{-1} $) & ($M _{\odot} $ yr$ ^{-1} $) && ($M _{\odot} $ yr$ ^{-1} $)&  ($M _{\odot} $ yr$ ^{-1} $) & ($M _{\odot} $ yr$ ^{-1} $) &   ($M _{\odot} $ yr$ ^{-1} $)   \\
\hline 
NGC~160 & 1.41 $\pm$ 0.63 & 0.89 $\pm$ 0.42 & 1.05 $\pm$ 0.50 & & 1.18 $\pm$ 0.47 & 0.74 $\pm$ 0.24 & 0.74 $\pm$ 0.29 & 0.82 $\pm$ 0.26\\
NGC~262 & 2.24 $\pm$ 1.00 & 2.86 $\pm$ 1.23 & 2.94 $\pm$ 1.26 & & 2.79 $\pm$ 1.12 & 2.77 $\pm$ 0.93 & 3.28 $\pm$ 1.37 & 3.10 $\pm$ 1.05\\
NGC~4698 & 0.12 $\pm$ 0.06 & 0.05 $\pm$ 0.03 & 0.11 $\pm$ 0.06 & & 0.15 $\pm$ 0.06 & 0.10 $\pm$ 0.03 & 0.08 $\pm$ 0.04 & 0.11 $\pm$ 0.04\\
NGC~5173 & 0.58 $\pm$ 0.26 & 0.45 $\pm$ 0.20 & 0.60 $\pm$ 0.26 & & 0.67 $\pm$ 0.27 & 0.32 $\pm$ 0.11 & 0.22 $\pm$ 0.09 & 0.27 $\pm$ 0.09\\
NGC~5389 & 0.41 $\pm$ 0.18 & 0.07 $\pm$ 0.04 & 0.11 $\pm$ 0.07 & & 0.44 $\pm$ 0.18 & 0.32 $\pm$ 0.10 & 0.29 $\pm$ 0.12 & 0.29 $\pm$ 0.09\\
NGC~5982 & 2.71 $\pm$ 1.21 & 0.09 $\pm$ 0.07 & 0.40 $\pm$ 0.23 & & 1.63 $\pm$ 0.65 & 0.68 $\pm$ 0.24 & 0.20 $\pm$ 0.08 & 0.37 $\pm$ 0.12\\
NGC~6962 & 4.11 $\pm$ 1.84 & 3.69 $\pm$ 1.63 & 4.24 $\pm$ 1.88 & & 2.64 $\pm$ 1.06 & 1.83 $\pm$ 0.63 & 2.14 $\pm$ 0.85 & 2.38 $\pm$ 0.78\\
NGC~7371 & 1.69 $\pm$ 0.76 & 2.33 $\pm$ 1.00 & 2.94 $\pm$ 1.26 & & 1.77 $\pm$ 0.71 & 1.03 $\pm$ 0.34 & 1.01 $\pm$ 0.39 & 1.18 $\pm$ 0.38\\
NGC~7787 & 2.40 $\pm$ 1.07 & 0.66 $\pm$ 0.30 & 1.01 $\pm$ 0.45 & & 2.59 $\pm$ 1.03 & 2.64 $\pm$ 1.11 & 2.88 $\pm$ 1.69 & 2.67 $\pm$ 1.35\\
PGC~065981 & 1.92 $\pm$ 0.86 & 3.57 $\pm$ 2.23 & 4.01 $\pm$ 2.63 & & 1.34 $\pm$ 0.54 & 1.94 $\pm$ 1.23 & 3.55 $\pm$ 2.01 & 3.83 $\pm$ 1.77\\
\hline  
\end{tabular}

\end{table*}

\subsubsection{Outer regions}
 \label{section432}


As discussed in the Introduction, we are interested in the outer parts of the galaxies. Because of the low angular resolution of the IR data, the expressions that use these to correct for dust attenuation can only be applied to the whole galaxy measurements, and not to specific regions, such as rings, the central parts, or spiral arms (an example of the problems induced by inaccurate estimates of local SFRs is given in \citealt{Calzetti2012}).

To quantify how much emission from the galaxy is located in the rings/spiral arms, we have computed the ratio of the observed luminosity of the feature and the observed luminosity of the complete galaxy, shown in Table \ref{percentages}. We have not included NGC~5982 in the table, as all the luminosity comes from the central part, contrary to what happens in the other elliptical of the sample, NGC~5173, with its emission in the form of clumps or incomplete spiral arms.

\begin{table*}
\caption{Ratios of the luminosities (not corrected for internal dust absorption) of the whole galaxy and of the outer feature. Column I) Galaxy name. Column II) Outer feature of the galaxy. Column III) Ratio of the observed luminosity of the feature and the luminosity of the whole galaxy in H$\alpha$, written $ L_{\rm H \alpha ,feat}/L_{\rm H \alpha ,tot}$. Columns IV and V) As column III, but for FUV and NUV, respectively.}
 \label{percentages}
\centering
\begin{tabular}{|c|c|c|c|c|c|c|c|}
\hline
Galaxy name & Feature & $ L_{\rm H \alpha ,feat} $ &  & $ L_{\rm FUV,feat} $& & $ L_{\rm NUV,feat} $ \\ \cline{3-3} \cline{5-5} \cline{7-7}
  &   & $ L_{\rm H \alpha ,tot} $& & $ L_{\rm FUV,tot} $& & $ L_{\rm NUV,tot} $  \\
\hline
NGC~160    & Ring   & 0.95&& 0.86& & 0.78 \\
 NGC~262    & Arms   & 0.03&& 0.71& & 0.60 \\
 NGC~4698   & Ring 1 & 0.20&& 0.40& & 0.36 \\
 NGC~4698   & Ring 2 & 0.80&& 0.60& & 0.64 \\
 NGC~5173   & Arms   & 0.54&& 0.59& & 0.60 \\
 NGC~5389   & Ring   & 0.57&& 0.56& & 0.45 \\
 NGC~6962   & Arms   & 0.76&& 0.91& & 0.87 \\
 NGC~7371   & Ring   & 0.99&& 0.98& & 0.72 \\
 NGC~7787   & Ring   & 0.62&& 0.76& & 0.66 \\
 PGC~065981 & Arms   & 0.80&& 0.81& & 0.77 \\
\hline\end{tabular}
\end{table*} 

 In Table \ref{percentages} we can see that in most cases, the emission of the outer features is more than half of the total emission from the galaxy. In fact, for NGC~160, NGC~6962 and NGC~7371, the luminosity of the feature is more than 75 per cent of the total luminosity. Also, if we add the emission of all the rings in NGC~4698, the sum reaches almost 100 per cent of the emission of the galaxy. On the other hand, NGC~262 is different: while the ratio in the UV light is more than 50 per cent, the emission in H$\alpha$ is nearly zero, which means that almost all the emission of the galaxy comes from the centre, and not from the arms. The reason for this may be that NGC~262 is a Seyfert 2 galaxy, and therefore the nuclear emission dominates (the H$\alpha$ emission found in the central parts of the galaxies may not be due to star formation, but also to AGN activity). Note that the central emission is not taken into account in Table \ref{percentages}, just the emission from the features.


\section{Discussion}
 \label{section5}

\subsection{Morphology}

The first important finding of this work is the good correlation between the H$\alpha$ and the UV morphologies in our sample galaxies. In other words, there is a direct correspondence between the location of the UV emission on the outskirts of the galaxy, or even outside it, as identified in the \textit{GALEX} images, and the H{\sc ii} regions present in the H$\alpha$ images. Some galaxies have morphological classifications which include outer (pseudo-)rings [designated by (R) or (R') preceding the main type symbols], i.e., NGC~160, NGC~7731 and NGC~7787. The H$\alpha$ emission in these outer rings is easily traced. It is not as extended as the UV emission, and shows as knots with patchy emission.

The main overall conclusion is that the excess UV emission in the studied elliptical, S0, and early-type spiral galaxies is due to massive star formation, eliminating the need to rely on other explanations such as the UV upturn phenomenon.


\subsection{Star formation rates}

The second result  of this work concerns the derived SFRs. We find relatively modest SFRs after correction for internal dust absorption, and we find that these values coincide in H$\alpha$, FUV and NUV, within the uncertainties. From these results we can derive three direct implications.

The first one is a confirmation that the global SFRs should be derived including both direct light and dust-processed light, in other words, corrections to the observed H$\alpha$ and UV luminosities are necessary. Tables \ref{SFRs} and \ref{SFRs2} show by how much the SFRs change when applying corrections for dust attenuation, more so in the UV.
 
The second implication arises from the fact that the dust-corrected SFRs in H$\alpha$, FUV and NUV are very similar. This confirms that the star formation in these galaxies is due to a recent star forming period, with both the UV and H$\alpha$ emission coming from a population of young, hot stars. Our data demonstrate that the galaxies in the sample have recent massive star formation, and that the excess UV-light is likely not due to the UV-upturn phenomenon. Furthermore, the levels of star formation found in the ellipticals of the sample are above-average for that morphological type, while still well below $1\,M _{\odot} $ yr$ ^{-1} $.

Finally, the third implication is that these UV-peculiar ellipticals, S0s and ETSs form stars at a rate comparable to  late-type galaxies (e.g., \citealt{James2004}). Specifically, in seven of the ten galaxies in our sample, the derived SFR(H$\alpha$) is larger than 1 $M _{\odot} $ yr$ ^{-1} $. This result is in good agreement with previous studies (e.g., \citealt{Young1996}; \citealt{Usui1998}; \citealt{Hameed1999} or \citealt{Hameed2005}, see Sect. \ref{section1}). 

\subsubsection{Corrections for internal dust absorption}

In Section \ref{sectionSFRs} we noted why it is necessary to correct the luminosities for internal dust absorption. For the galaxies under consideration here, however, the methods proposed in the literature and used here have both advantages and drawbacks. First of all, due to the fact that they are expected to have less dust than spirals, one should be careful when adopting the same correction for all types of galaxies regardless of their morphological type or other characteristics. Furthermore, our sample galaxies span a wide range of stellar masses and SFRs, from truly quiescent galaxies to ones with high SFRs, which hinders the use of one single correction.

Our first attempt to correct both UV and H$\alpha$ data has been using the TIR luminosity to correct for internal dust absorption, using the equations presented in \citet{Kennicutt2009} and \citet{Hao2011}. In particular in systems with low specific SFR, a fraction of the IR light can come from dust heated by old population stars, which can lead to an overestimate of the corrected IR luminosity and consequently of the derived SFRs. As we do not know a priori whether a galaxy has a low specific SFR, we cannot blindly apply the equations presented in \citet{Kennicutt2009} and \citet{Hao2011} (our equations \ref{Lcorr1}, \ref{Lcorr2} and \ref{Lcorr3}). 

This "dust heating" problem leads us to prefer other methods for dust correction. In the case of the UV imaging, we use the relationship between TIR/FUV and UV attenuation proposed by \citet{Cortese2008}, which is independent of the age of the stellar populations and can be applied to systems with low specific SFRs.

For H$\alpha$, we thus use the recipes presented by \citet{Helmboldt2004}, relating the absolute \textit{R} magnitude with the H$\alpha$ absorption coefficient (our equation \ref{helmboldt}). Table \ref{extinction} shows that $<A(\rm{H \alpha})_{\rm TIR} >$=0.73 mag, lower than $<A({\rm H \alpha} ) _R >$=1.18 mag, but both methods are more reliable than adopting a single value of $A(\rm {H \alpha} )$=1.1 mag. (see Sect. \ref{sectionSFRs}). Our preferred corrected SFR values, along with those arrived at by using the TIR luminosity, are listed in Table \ref{SFRs2}.

\subsection{Origin of star formation}

One of the main results of the current study is that the origin of the UV emission detected beyond $R_{25}$ in  number of elliptical, S0, and ETS galaxies lies in massive star formation, in particular located in outer spiral arms and rings. But why do these galaxies have UV emission, and massive star formation beyond $R_{25}$? The rings may well be related to the presence of bars, but the small sample size (only five galaxies with rings, classified as SB [1], S\underline{A}B [1], SAB [1] and SA [2]) does not allow us to explore this further. 

\citet{Carter2011} studied the origin of the FUV excess in elliptical and S0 galaxies, and suggested that this excess can be moderately strong in galaxies which show evidence of recent minor mergers, for instance the shell elliptical NGC 5982 \citep{Sikkema2007}. In this galaxy, we find moderate levels of star formation, and agree that this may be due to a recent merging event. Furthermore, NGC~5982 was classified as a LINER (\citealt{Ho1997}) and the H$\alpha$ emission from the centre may not be due to star formation, but to the non-stellar nuclear activity.

NGC~5173 is a remarkable case. This E galaxy appears as elliptical in the $R$-band image, as well as in other IR images (see NED). However, the galaxy shows a lot of extra emission in both the H$\alpha$ and \textit{GALEX} images (see Fig. \ref{fig2}). The morphology of this emission seems to be in the form of incomplete arms. NGC~5173 is in a group of galaxies \citep{Mahtessian1998}, with NGC~5169 at a projected distance of 64.8 kpc and a difference in velocity of 17 km/s. The interaction with this companion might well be related to the enhanced star formation.

The (FUV$-$NUV) colour can be understood as a discriminator between elliptical/lenticular galaxies and spiral galaxies: the former are expected to have FUV$-$NUV$>0.9$ (\citealt{GildePaz2007}). We find (last column of Table \ref{photometry}) that only NGC~4698 and NGC~5982 have FUV$-$NUV$>0.9$. The other galaxies have FUV$-$NUV$<0.9$, confirming that they present an unexpected behaviour for their types, and that they behave more as star-forming galaxies than as more standard, red, and inactive galaxies. This colour also confirms that NGC~4698 behaves as an elliptical rather than a spiral galaxy, which matches better the latest morphological classification by R. J. Buta et al. 2013 (S0/a-E2) than the earlier one in the RC3 (Sab).

Our limited study thus does not let us conclude much on the origin of the extended massive star formation, beyond the obvious statements that a certain fraction of galaxies has star-forming rings and spirals beyond $R_{25}$, and that interactions and (minor) mergers may stimulate this star formation. 


\section{Conclusions}
 \label{section6}

We investigate how many elliptical, S0, and ETSs show excess UV emission from beyond their optical extent ($R_{25}$), and have obtained deep H$\alpha$ imaging of ten of them, with the goal of understanding its nature. From our analysis, we conclude the following:

\begin{enumerate}

\item From {\it GALEX} images of 1899 elliptical, S0, and ETS galaxies, we find that 28\% of them show excess UV emission beyond $R_{25}$. This fraction rises from 18\% of the elliptical galaxies to 28\% for the S0s and 45\% of the ETSs.

\item After obtaining deep H$\alpha$ imaging of a subsample of 10 of these UV-excess galaxies, we detect H$\alpha$ emission at the same location as the UV emission in the outer parts of all but one of the sample. There is good correspondence between the H$\alpha$ and UV morphology of the outer features of the galaxies.

\item The above implies, at least in the limited number of cases studied in detail here, that the excess UV emission detected with \textit{GALEX} originates in standard massive star formation processes, in most cases occurring in outer features such as spiral arms or outer rings. From the presence of the H$\alpha$ emission, we conclude that there is no lack of the most massive stars populating the initial mass function, as postulated in other regimes \citep{Lee2011}. There is no need to invoke other explanations for the excess UV emission, such as the UV upturn phenomenon.

\item After correcting the observed H$\alpha$, FUV and NUV luminosities of the galaxies for internal dust absorption, there is a good agreement in the derived SFRs, with values of a few tenths to a few $M_{\odot}$\,yr$^{-1}$, comparable to SFR values found in late-type galaxies. The two elliptical galaxies in our sample have SFRs below $1\,M_{\odot}$\,yr$^{-1} $.

\end{enumerate}

\section*{Acknowledgments}

We would like to thank Barry Madore for his comments and suggestions during the preparation of this manuscript, Mauricio Cisternas and Mark Seibert for their useful comments, and Jos\'e Ram\'on S\'anchez-Gallego for his assistance with the reduction and calibration of the H$\alpha$ images. SE-F, JER and LMZH would like to thank the Carnegie Observatories for hospitality and facility support. JHK acknowledges financial support to the DAGAL network from the People Programme (Marie Curie Actions) of the European Union's Seventh Framework Programme FP7/2007-2013/ under REA grant agreement number PITN-GA-2011-289313. EANMNV thanks the IAC for a summer studentship and financial support. JER and LMZH would like to thank the GLGA and the Rose Hills Foundation for financial support. GLGA funding was from NASA grant NXOHAKS6G. JER acknowledges partial support from the University of Wisconsin College of Letters \& Science. The Institute for Gravitation and the Cosmos is supported by the Eberly College of Science and the Office of the Senior Vice President for Research at the Pennsylvania State University. Based on observations made with the NOT operated on the island of La Palma, in the Spanish Observatorio del Roque de Los Muchachos of the Instituto de Astrof\'isica de Canarias. The data presented here were obtained in part with ALFOSC, which is provided by the Instituto de Astrof\'isica de Andalucia under a joint agreement with the University of Copenhagen and NOTSA. Some of the data presented in this paper were obtained from the Mikulski Archive for Space Telescopes (MAST). STScI is operated by the Association of Universities for Research in Astronomy, Inc., under NASA contract NAS5-26555. Support for MAST for non-\textit{HST} data is provided by the NASA Office of Space Science via grant NNX09AF08G and by other grants and contracts. This research has made use of the NASA/IPAC Extragalactic Database (NED) which is operated by JPL, Caltech, under contract with NASA.


\bibliographystyle{mn2e}
\bibliography{references}

\bsp

\label{lastpage}

\end{document}